\RequirePackage{lineno}
\documentclass[aps,pre,twocolumn,superscriptaddress,showpacs, lineno]{revtex4}
\usepackage{amssymb}
\usepackage{epsfig}
\usepackage{amsmath}
\usepackage{times}
\usepackage{color}
\usepackage{lipsum}
\usepackage{ulem}
\usepackage{sidecap}

\newcommand{\red}{\textcolor[rgb]{0.00,0.00,0.00}}

\begin{document}
\title{Spiral wave chimeras in locally coupled oscillator systems}
\author{Bing-Wei Li}
\email[Corresponding author. Email address: ]{bwli@hznu.edu.cn}
\affiliation{Department of Physics, Hangzhou Normal University,
Hangzhou 310036, China}
\affiliation{Department of Physics and Astronomy, Ghent University, Krijgslaan 281, 9000 Gent, Belgium}
\author{Hans Dierckx}
\affiliation{Department of Physics and Astronomy, Ghent University, Krijgslaan 281, 9000 Gent, Belgium}

\begin{abstract} 
The recently discovered chimera state involves the coexistence of synchronized and desynchronized states for a group of identical oscillators. This fascinating chimera state has until now been found \red{only} in non-local or globally coupled  \red{oscillator} systems. In this work, we \red{for the first time show numerical evidence of}  the existence of spiral wave chimeras in reaction-diffusion systems \red{where each element} is {\it locally} coupled by diffusion. This spiral wave chimera rotates inwardly, \red{i.e., coherent waves propagate toward the phase randomized core}. A continuous transition from spiral waves \red{with smooth core} to spiral wave chimeras is found as we change the local dynamics of the system. Our findings on the spiral wave chimera in locally coupled oscillator systems largely improve our understanding of the chimera state and suggest that spiral chimera states may be found in natural systems which can be modeled by a set of oscillators indirectly coupled by a diffusive environment.
\end{abstract}
\date{\today }
 \maketitle

\section{Introduction}
Collective behavior, which occurs commonly in physical, chemical and biological systems, has been a subject of continued interest in nonlinear science over the last decades \cite{winfree,pikovsky:book,Strogatz:book,kuramoto}. In neural and biological systems, a typical collective behavior observed is the coherent motion of oscillators. This phenomenon of synchronization has been widely regarded as having important implications to the function and performance of those systems \cite{goldbeter,glass}. For instance, asynchronous contraction of the heart the may be triggered by electrical spiral waves, which eventually leads to heart dysfunction \cite{gray}.

Recently, much attention has been paid to a particular hybrid state in which an ensemble of identical oscillators with identical coupling spontaneously degenerate to one group with synchronization (coherent) and the other group with desynchronization (incoherent) \cite{kuramoto02,shimapre04,abrams04,hagerstromnat12,tinsleynat12,nkomoprl13,martenspns13,sethiaprl14,martensprl10,guprl13,yeldesbayprl14}. This fascinating counterintuitive state was first discovered by Kuramoto and co-workers \cite{kuramoto02}, and named ``chimera state" by Strogatz \cite{abrams04}. The experimental confirmation of the existence of chimera states has been reported independently almost at the same time in diverse systems such as coupled maps \cite{hagerstromnat12}, chemical oscillators \cite{tinsleynat12,nkomoprl13} and mechanical pendulums \cite{martenspns13}. In two-dimensional systems, chimera states take the form of spiral waves \cite{shimapre04,martensprl10,guprl13,wangjcp}, called spiral wave chimeras with phase-locking oscillators in the spiral arm but a phase-randomized spiral core \cite{shimapre04}. An analytical solution for a spiral wave chimera was further demonstrated by Strogatz {\it et al}. using a Kuramoto-type phase equation with nonlocal coupling \cite{martensprl10}. Recently, the spiral wave chimera state has been reported experimentally in chemical oscillators \cite{nkomoprl13} and numerically in complex and even chaotic oscillators \cite{guprl13} where nonlocal coupling is introduced.

It was initially believed that \red{the chimera state arises} from so-called nonlocal coupling, i.e. each oscillator in the system will be affected {\it instantaneously} by a group of oscillators within certain interacted range. Therefore, nonlocal coupling is intermediate between the cases of local and global coupling. However, some recent works show that the non-locality conditions for occurrence of the chimera state can be further relaxed \cite{yeldesbayprl14, sethiaprl14}. For example, Sethia {\it et al}. have demonstrated a generalized chimera state called ``amplitude mediated chimera state'' in a population of oscillators even in the case of global coupling \red{rather than the nonlocal coupling} \cite{sethiaprl14}.

It is worth pointing out, on one hand, that in the seminal work of Kuramoto {\it et al}., a key assumption to observe spiral wave chimeras is that the third component of reaction-diffusion (RD) changes so fast that it can be eliminated adiabatically \cite{shimapre04}. In this way, the three-component RD system is essentially reduced to an effective two-component one with an extra term represented by a nonlocal coupling. On the other hand, to observe chimera state in experiments, the realization of nonlocal coupling strongly relies on a computer \cite{hagerstromnat12,tinsleynat12,nkomoprl13,martenspns13}. Therefore, it remains an open question whether spiral wave chimeras exist in more natural systems. In such systems, \red {the time scale of each component in the system may be comparable} and the coupling will be mediated by a natural law such as diffusion. Further more, spiral waves can rotate outwardly or inwardly \cite{vanagsci}, however all the spiral wave chimeras reported previously are outward. There is very little information on the existence of inwardly rotating spiral wave chimera.

\red{In this work, we report the first numerical evidence of the existence of spiral wave chimeras in three-component RD system where each oscillatory element is locally coupled by diffusion, i.e., by nearest neighbor interaction. Differing from previously observed spiral wave chimeras in nonlocally coupled systems, our study shows remarkable novel properties of spiral wave chimeras: first the observed spiral wave chimeras rotate inwardly, i.e., a coherent wave propagates toward the phase-randomized center. Second, the non-diffusing components in the RD system exhibit chimera properties, while the diffusing components show a coherent spiral structure. A continuous transition from spiral wave with a smooth core to spiral wave chimeras is also identified. Our findings in RD systems show that the occurrence of the chimera states does not require nonlocal or global coupling and therefore provide key hints to explore the chimera state in the natural world.}

\section{Reaction-Diffusion Model and Methods}
\textit{Model description. }
Our starting point is a three-component RD system in two spatial dimensions \cite{alonsojcp11,nicolapre02}:
\begin{eqnarray}
\partial_{t} X &=& \phi(aX -\alpha X^{3}-bY -cZ)+D_{X}\nabla^{2} X,\nonumber 
\\
\partial_{t} Y &=& \phi\epsilon_{1}(X - Y), 
\label{rd} \\
\tau\partial_{t} Z &=& \phi\epsilon_{2}(X - Z) + D_{Z}\nabla^{2} Z. \nonumber 
\end{eqnarray}
These equations describe the evolution of concentrations of chemical reactants $X$, $Y$ and $Z$, where $X$ is an activator and $Y$ and $Z$ are inhibitors. $D_{X}$ and $D_{Z}$ denote the diffusion coefficients of chemical species of $X$ and $Z$, respectively. The dimensionless parameter $\tau$ represents the characteristic time scale of the system variable $Z$, and will be assigned a finite non-zero value. \red{In principle, the parameter $\phi$ can be absorbed to $a$, $b$, $c$, $\alpha$, $\epsilon_{1}$ and $\epsilon_{2}$ which represent other parameters, but we still write it explicitly here to keep the same form as in Ref. \cite{alonsojcp11}.} This three-component RD system actually is an extension of the two-component FitzHugh-Nagumo model by coupling to a third variable $Z$, which was proposed to study pattern formation in the Belousov-Zhabotinsky (BZ) reaction dispersed in water droplets of a water-in-oil
aerosol OT (AOT) microemulsion system (BZ-AOT system) \cite{alonsojcp11} and to model spot dynamics in gas discharges \cite{schenkprl97}.

This study will differ from previous work on pattern formation in RD systems in two aspects. The first modification is related to the ratio of the diffusion coefficients $\delta = D_{X}/D_{Z}$. Traditionally $\delta \geq 1$, but in our study we take $\delta \ll 1$ or $\delta = 0$. Therefore, we consider Eqs. \eqref{rd} from a dynamical point as a model describing a large number of oscillators that communicate with each other via a diffusive environment. Such kind of models may be related to various systems such as bacteria, yeast cells and chemical oscillators. The second important thing is on the time scale of $\tau$. Our second modification is to omit the key assumption that the third variable $Z$ changes fast, i.e., $\tau \rightarrow 0$ \cite{shimapre04, nicolapre02}. Hence, in our case the above three-component RD system can no longer be reduced a two-component system with traditional nonlocal coupling.

\textit{Numerical methods. }
We numerically integrated Eqs. \eqref{rd} in time using the explicit forward Euler method with spatial step $\Delta x = \Delta y = 0.2$ and time step $\Delta t = (\Delta x)^{2}/(5D_{Z})$, in a grid of $N_{x}=512$ by $N_{y}=512$ oscillators. We use a five-point stencil to evaluate the Laplacian term in \eqref{rd}. Note that the diffusion term in \eqref{rd} is effectively implemented as the interaction of each oscillator with its four nearest neighbours  on a rectangular lattice. In this paper, we focus on the dynamics of system as we vary the model parameters $a$, $\epsilon_{2}$, with fixed value of $\phi = 0.62$, $b = 3.0$, $c=3.5$, $\epsilon_{1} =1.0$, $\alpha = 4.0/3.0$, $\tau=1.0$. In the case of $\delta =0$, changing the diffusion coefficient $D_{Z}$ is equivalently to rescale the size of the system, and thus we keep the same value $D_{Z}=0.5$ through the work.

\begin{figure}[tbp]
\centering
\includegraphics[bb=0pt 0pt 254pt 240pt,clip,scale=1.0]{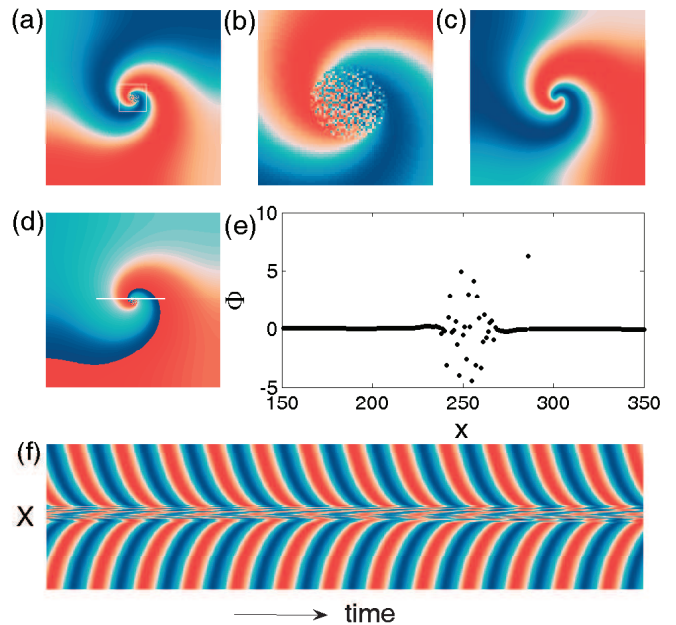} %
\caption{(color online) Spiral wave chimeras in a three-component RD system \eqref{rd} with $a=1.50$ and $\epsilon_{2}=0.15$ . (a) denotes snapshot of the variable $X$; (b) is the magnification the square region in (a); (c) represents the variable of $Z$; (d) is the snapshot of phase defined $\tan \theta(\vec{r}) = Y(\vec{r})/X(\vec{r})$ and (e) is a cross section along the line in (d) by defining $\Phi = \theta_{I+1,J}-\theta_{I,J}$. It is noted that spiral core is incoherent for the variable $X$, while it is coherent for the variable $Z$. (f) is the spatiotemporal pattern of $X$ along the middle horizontal line of (a) showing  wave propagation towards the spiral chimera core. In panels (a-d), the wave pattern rotates clockwise.}
\end{figure}

\begin{figure}[tbp]
    \includegraphics[bb=0pt 0pt 226pt 233pt,clip,scale=1.0]{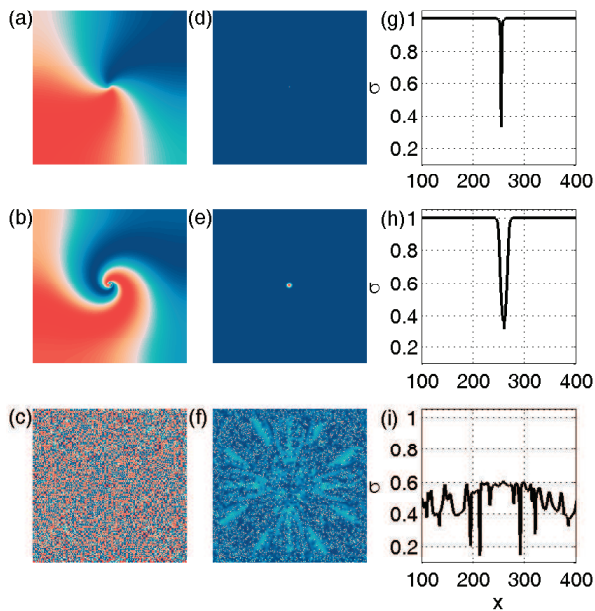}
    \caption{(color online) Transition of chimera spiral waves as a function of $a$.
In the case of small $a=0.90$, a coherent spiral wave is observed (a) and it is
gradually changed to a spiral wave chimera (b) as we increase the system parameter $a=1.40$. Further increasing $a$ led to the formation of completely disorder state (c) where $a=1.70$. We here also show the corresponding time averaged order parameter $\sigma$(refer to (d-f)) and the order parameter but along the center of medium (g-i). Except of $a$, other parameters are taken as in Fig. 1.}
\end{figure}

\textit{Order parameter. }
To analyze the state of the coupled oscillators, we introduce the oscillator phase $\theta$ by $\tan \theta(\vec{r}) = Y(\vec{r})/X(\vec{r})$. Since several observed states will be incoherent at the level of the simulation grid, we consider the domain as a discrete set of oscillators at positions $x=I\Delta x, y=J\Delta y$ with phase $\theta_{I,J}=\arctan (Y_{I,J}/X_{I,J})$.

To quantitatively study the size of the region with incoherent oscillations, we define the spatiotemporal order parameter
\begin{equation}
\sigma_{I,J}(t)=\frac{1}{(2m+1)}\left|\sum_{<I,J>}\exp[i\theta_{I,J}(t)]\right|
\end{equation}
where $\theta_{I,J}$ is the oscillator phase as defined above. The notation $\left<I,J\right>$ denotes the set of nearest neighbors include itself and $1/(2m+1)$ is a normalization factor where $m$ denotes the number of oscillators with the nearest coupling along a given spatial dimension. In the present case, we set $m=2$, since we know the diffusion term in Eqs. \eqref{rd} was computed using a 5-point discrete Laplacian. Finally, the time-averaged order parameter was computed as $\overline{\sigma}_{I,J}= \lim_{\Delta T\rightarrow \infty}  \frac{1}{\Delta T} \int_{t_{0}}^{t_{0}+\Delta T} \sigma_{I,J}(t') dt'$, where $\Delta T$ is a  time interval, e.g, $\Delta T=20000$ in our simulations. This quantity was used to estimate the radius of the incoherent state $R=(R_{x}+R_{y})/2$ where $R_{x}$ ($R_{y}$) denotes maxim distance between the grid points along the $x$ ($y$) line when $\overline{\sigma}_{I,J}\leq 0.9$.

\section{Results}
\textit{Existence of chimera state in a RD system with inhibitor diffusion only ($\delta=0$). }
Figure 1 demonstrates the existence of a spiral wave chimera state in the locally coupled RD system \eqref{rd}. The snapshot of variable $X$ and its magnification around the core are shown in Fig. 1(a) and (b), respectively at time $t=10^{5}$ after the initiation of the spiral wave. Note that the $X$-variable around the spiral core is discontinuous while it is smooth far from the core, which is similar to previously reported spiral waves chimeras in nonlocal systems.  This property is also true for the other non-diffusing variable $Y$. (Figure not shown).

However, in contrast to $X$ and $Y$, for the diffusing inhibitor $Z$, the whole spiral pattern is smooth even in the region closed to the spiral core as shown in Fig. 1(c). This property is different from the spiral wave chimeras found in nonlocally coupled oscillator systems where all of variables describing the oscillator demonstrate a similar chimera characteristic.

The corresponding phase distribution of $\theta_{I,J}$ and the dependency of $\Phi_{I,J}=\theta_{I,J}-\theta_{I+1,J}$ on the position along a horizontal cross section through the center of the medium ($J=N_{y}/2$) is illustrated in Fig. 1(d-e). As one notes that close to spiral center, the quantity $\Phi_{I,J}$ are nonzero, which means the phases of the oscillators around the spiral core are discontinuous or incoherent; while far way from it, the phases becomes continuous or coherent as indicated by the zero quantity of $\Phi_{I,J}$. This behavior is the defining property of a chimera state for a group of identical oscillators: some oscillate in a coherent way while a localized subgroup oscillate incoherently \cite{kuramoto02}.

It is worth pointing out that the spiral wave chimera shown in Fig. 1 rotates inwardly, i.e. one sees the coherent waves propagate toward the phase-randomized core. This can also be seen from the spatiotemporal pattern for $X$ variable along a horizontal cross section through the center of the medium ($J=N_{y}/2$) shown in Fig. 1(f). The inwardly rotating spiral waves, also called antispiral waves, have been reported 15 years ago \cite{vanagsci}. However, to our best knowledge, this is the first time to report such inwardly rotating wave chimeras in a RD system.

\textit{Impact of local dynamics on the formation of spiral wave chimera.}
To get more insights to the formation and robustness of the spiral wave chimera, we investigate its behavior under a sweep of the parameter $a$. \red{From \cite{alonsojcp11}, we recall that the system \eqref{rd} changes from a stationary state to an oscillatory state via a Hopf bifurcation when the value of the parameter $a$ is increased}. For $a \leq 1$, we observe spiral waves (inward) with a smooth core. From, $a \leq 1.1$ onwards, spiral chimera states are found, whose core radius increases with increasing $a$; compare panels (a) and (b) from Fig. 2. The incoherent region of the chimera state monotonically grows with increasing $a$, until the incoherent state fully occupies the domain. This completely incoherent state is already reached in  Fig. 2(c) for $a=1.7$. \red{We have found continuous transition from normal spiral waves to spiral wave chimeras by controlling the parameter $a$.}

In Fig. 2(d-f), we show the time-averaged order parameter for different values of $a$. It is noted that in the incoherent chimera core, $\overline{\sigma}_{I,J}$ is smaller than one whereas in the outer region $\overline{\sigma}_{I,J}\approx 1$ due to the continuity of phase. Therefore, in the case of spiral waves, the region with $\overline{\sigma}_{I,J}\approx 1$ is extremely small, see Fig. 2(d). On the contrary, for the turbulent state, $\overline{\sigma}_{I,J} \ll 1$ everywhere. These statements can also be seen clearly from Fig. 2(g-i), which show the local order parameter $\overline{\sigma}_{I,J}$ along the line $J=N_{y}/2$.

The dependency of the chimera core radius on the parameter $a$ is shown in Fig. 3. When $0.90 \leq a\leq 1.0$, the radius $R \approx \Delta x$ and thus in such a case, a classical spiral wave emerges. However, when $a$ lies between $1.1$ and $1.5$, the radius of incoherent spiral core is clearly finite, corresponding to the spiral wave chimera state. \red{Further, we observe a fast transition from spiral waves chimera to completely turbulent state beyond $a = 1.5$. }

\begin{figure}[btp]
\includegraphics[bb=66pt 238pt 515pt 455pt,clip,scale=0.55]{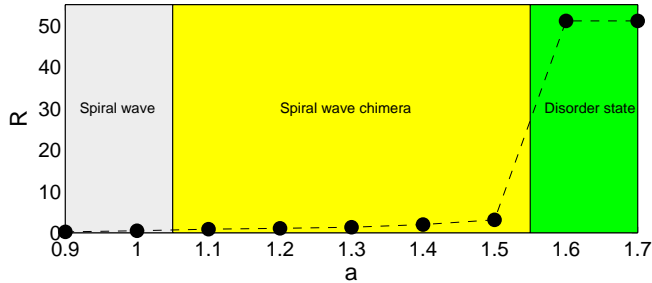} %
\caption{(color online) Radius of the incoherent chimera core as a function of model parameter $a$, for $\epsilon_2= 0.15$.}
\end{figure}

To give the global picture about how the local dynamics parameter affects the behaviors of the spiral wave chimeras, we further identified the occurrence of the chimera state in a wide range of the $a$-$\epsilon_2$ parameter space. The results are summarized in Fig. 4. In this figure, full circles represent the stable spiral wave chimeras, the full triangles and squares denote the spiral waves and incoherent state. The cross means that oscillations are not sustained. In the parameter space, the separation line between spiral and spiral wave chimera is almost vertical, which means that $a$ plays the key role in determining such state, while $\epsilon_{2}$ has a more important role in determining whether spiral chimeras or turbulent states appear.


\begin{figure}[t]
\includegraphics[bb=38pt 202pt 383pt 531pt,clip,scale=0.65]{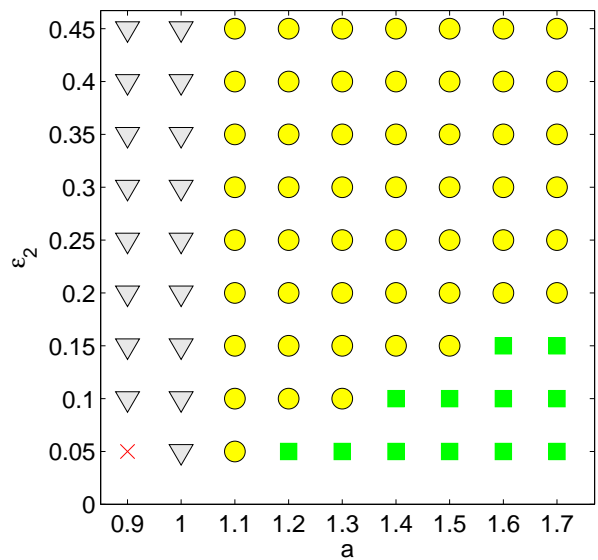} %
\caption{(color online) Phase diagram for the spiral wave chimera state in $a-\epsilon_{2}$ parameter space showing occurrence of spiral waves($\Delta$), fully incoherent state ($\square$), spiral wave chimera states ($\bullet$) and a stable resting state ($\times$). All the pattern were observed at $t=10^{5}$ after the initiation of the spiral wave.}
\end{figure}

\textit{Spiral wave chimeras with finite activator diffusion ($\delta$). }
Till now, we only consider the RD systems where only the diffusion of inhibitor, $Z$, is presented, i.e., $\delta = 0$. A natural question is: how such kind of spiral wave chimeras would change when we increase $\delta$ from zero to a finite value? Typical results are presented in Fig. 5. For $\epsilon_{2}=0.45$, spiral wave chimeras survive as we increase to $\delta=10^{-3}$, see Fig. 5(a) and
magnification of the core region 5(d). The core region becomes smooth but somehow spilt as we increase to $\delta=10^{-2}$, see Fig. 5(b) and 5(e). Further increasing $\delta=10^{-1}$ finally leads to the appearance of the spiral wave with smooth core as illustrated in Fig. 5(c) and 5(f). These results are within our expectation as the diffusion coupling of the activator has a clear smoothing effect.


%

\begin{figure}[tbp]
\includegraphics[bb=0pt 0pt 252pt 159pt,clip,scale=1.0]{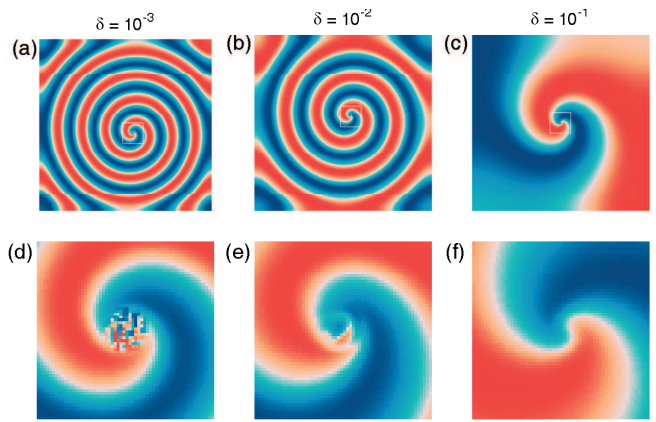} %
\caption{(color online) The impact of finite $\delta$ on spiral wave chimera. (a) $\delta=10^{-3}$ Spiral wave chimeras; (b) $\delta=10^{-2}$ Spiral wave with split core; (c) $\delta=10^{-1}$ Spiral wave with smooth core. (d-e) are the corresponding core region as indicated by the white square region in (a-c). $\epsilon_{2} = 0.45$, $a=1.70$.  }
\end{figure}

\section{Discussion}

In this study, we have found states in locally coupled RD systems that are highly similar to the chimera states found in non-locally coupled oscillator systems \cite{shimapre04}. We have identified this spiral wave chimera state for a broad parameter range in our model that is based on a chemical system.  Our results are not limited to the particular RD system \eqref{rd}, when it is are linearly coupled to the a third variable that diffuses. Since no parameter tuning was involved in our numerical experiments, we believe that the robust nature of spiral wave chimeras found in locally coupled systems indicates that spiral wave chimeras may be a frequent self-organized spatiotemporal pattern in complex systems, which may arise from weak coupling of a set of oscillators by a diffusing variable. Therefore, we expect that natural chimera states will be found in oscillatory biological or chemical systems.

Previous works on chimera states were focused on nonlocal or global coupling. In both cases, each element in the system was treated as a discrete oscillator. Whether chimera states exist in a spatially extended continuous media remains an open question. In a recent note \cite{laingarxiv}, one-dimensional chimera states in networks of pure local coupling were recently reported. However, this author did not consider spiral chimera states.

In our work, we started from the spatial discretization of a Laplacian operator in Eq. \eqref{rd} on a Cartesian grid, and found that the resulting wave became incoherent at the level of the grid scale. Furthermore, we have verified that using a finer discretization, the chimera state still persists. Thus, in locally coupled systems, the elusive chimera state breaks the continuum representation of the medium and forces one to consider the domain as a set of individual oscillators. Therefore, future analytical approaches will need to match the continuum dynamics in the outer region with discrete non-linear dynamics in the spiral core region.

\red{
It is noted that although we take the form of local coupling between the $Z$ variable, in the limit case of $\delta = 0$, such kind of local coupling of the inhibitor $Z$ may give rise to the nonlocal effect due to the in absence of diffusion of activator $X$. However, these nonlocal effects, which largely differs from the traditional nonlocal coupling (where each oscillator in the system will be affected {\it instantaneously} by a group of oscillators within certain interacted range), are not only spatial but also temporal, because of the finite $\tau$. The systems with the local coupling that nevertheless demonstrate spatiotemporally nonlocal effects have been largely overlooked in the last decades. However, such systems may be very common in biological systems and thus deserve further investigation in the future.
}

\red{
There may be various reasons for not observing chimeras in natural experimental settings before. From our study we note two necessary conditions: first, the local dynamics of each element needs to be oscillatory. Secondly, one requires close-to-vanishing spatial coupling of the observed variable.
}

\red{
Finally, we note the essential property of the system (1) used in our study is that each oscillator (described by $X$ and $Y$) is coupled indirectly via a nonuniform dynamical environment which described by the variable $Z$. The systems with such property may denote a broad class of the systems such as chemical oscillators e.g., BZ-AOT \cite{vanagsci} and BZ particles imersed in the solutions \cite{Taylorsci09}, engineered gene network \cite{Daninonat10}, yeast cells \cite{shutzBJ11}.  From the results in this study, we expect that natural chimera states can be found in many oscillatory biological or chemical systems.
}

\section{Conclusions}

In summary, we have shown that (inward) spiral wave chimeras do exist in spatially extended oscillatory media where only nearest-nearest interaction between the element is present. In such system, the nondiffusing components appear as a spiral wave chimera, while the diffusing variables show a coherent spiral wave structure.

A continuous transition from coherent spiral waves to spiral waves chimeras is observed as we increase the model parameter $a$ that controls the Hopf bifurcation. A phase diagram for spiral wave chimeras is identified
in the wide parameter space. We further discussed the smooth effects on the spiral core by the diffusion of the activator $X$. Our results on the spiral wave chimera in locally coupled oscillator systems largely improves our understanding of the chimera state and provides indications that the chimera state can be found in natural systems such as biological and chemical systems where each oscillating element communicates via a diffusive dynamical environment.

\begin{center}
\textbf{ACKNOWLEDGMENT}
\end{center}
This work was supported by the National Natural Science Foundation of China under Grants No.
11205039. H.D. received funding from FWO Flanders.

\end{document}